\documentstyle[12pt,aaspp4]{article}

\begin{document}

\def\kms{km~s$^{-1}$}
\def\msun{$M_{\odot}$}
\def\rsun{$R_{\odot}$}
\def\lsun{$L_{\odot}$}
\def\halpha{H$\alpha$}
\def\hbeta{H$\beta$}
\def\hgama{H$\gamma$}
\def\hdelta{H$\delta$}
\def\Teff{T$_{eff}$}
\def\logg{$log_g$}

\tighten


\title{Discovery of a Magnetic DZ White Dwarf with Zeeman-Split Lines
of Heavy Elements\altaffilmark{1}}

\author{I. Neill Reid}
\affil{Dept. of Physics \& Astronomy}
\affil{University of Pennsylvania}
\affil{209 S. 33rd St.}
\affil{Philadelphia, PA 19104-6396}
\affil{inr@herschel.physics.upenn.edu}

\author{and}

\author{James Liebert and Gary D. Schmidt}
\affil{Steward Observatory}
\affil{University of Arizona, Tucson AZ 85721}
\affil{jliebert,gschmidt@as.arizona.edu}

\begin{abstract}

A spectroscopic survey of unstudied Luyten Half Second
proper motion stars has resulted in the discovery of an
unusual new magnetic white dwarf. LHS~2534 proves to be
the first magnetic DZ, showing Zeeman-split \ion{Na}{1} and \ion{Mg}{1}
components, as well as \ion{Ca}{1} and \ion{Ca}{2} lines for which Zeeman
components are blended. The \ion{Na}{1} splittings result in a mean surface
field strength estimate of 1.92~MG.  Apart from the magnetic field, LHS~2534
is one of the most heavily-blanketed and coolest DZ white dwarfs at
$T_{\rm eff}\sim6,000$~K.

\end{abstract}


\altaffiltext{1}{The data presented here were obtained at the W.M.
Keck Observatory, which is operated as a scientific partnership among the
California Institute of Technology, the University of California, and the
National Aeronautics and Space Administration. The Observatory was made
possible by the generous financial support of the W.M. Keck Foundation.}

\section{Introduction}

Magnetic white dwarfs comprise $\sim$5\% of all white dwarfs and have surface
field strengths in the range $\sim$3$\times$10$^4$ to 10$^9$ gauss (G).  The
65 isolated (non-interacting binary) cases known at the time of the review of
Wickramasinghe \& Ferrario (2000) cover most of the white dwarf spectral types
(eg. DA, DB, DQ, DC), but have up to now not included any DZ stars which show
lines of heavy elements like Ca, Mg, Na and Fe.  This has restricted their use
as astrophysical laboratories of the effects of strong magnetic fields on the
light elements hydrogen, helium and molecular carbon.  The subject of this
paper is the discovery of the first magnetic DZ object, identified in the
course of routine spectral classification of cool stars from the Luyten
Half Second (1979, LHS) proper motion catalog.  We present in $\S$2 the
spectrum of this object, LHS~2534 (WD~1221-023, using the notation of McCook
\& Sion 1999). This dwarf offers the first empirical
data in an astrophysical setting of the Zeeman effect on neutral Na, Mg, and
both ionized and neutral Ca.

\section{The LHS~2534 Spectrum}

The optical spectrum of the new magnetic white dwarf was obtained on 8 February 1998
using the Low Resolution Imaging Spectrograph (LRIS, Oke et al. 1994) at the
10m W. M. Keck Observatory (Keck-II) on Mauna Kea, Hawaii.  These
observations were made as part of a service observing
request.  The 300 g/mm
grating blazed at 5000~\AA\ was used with a one arc second slit to obtain
spectra of 6\AA\ resolution covering $3800-7600$~\AA. A 
single 900-second exposure was obtained, and the data are plotted in Figure~1.

In addition to being magnetic, LHS~2534 is one of the most
heavily-blanketed of the known cool DZ white dwarfs.  The temperature is
evidently not too different from that of the Sun, as the strongest features --
\ion{Ca}{2} 3933, 3968~\AA, \ion{Ca}{1} 4226~\AA, \ion{Mg}{1} 5175~\AA, and
\ion{Na}{1} 5892AA\ -- are also among the strongest in the optical spectrum of
the Sun.  Hydrogen, especially H$\alpha$, is not detected, so one may conclude
the star has a helium-dominated atmosphere, like most DZ stars.

Monochromatic magnitudes for many cool white dwarfs were measured using
the Palomar Multichannel Spectrophotometer colors and published by Oke
(1974) and Greenstein (1976); the colors b(4255~\AA), g(4717~\AA), 
v(5405~\AA), r(6944~\AA) and i(8000~\AA) overlap the wavelength range of
these spectra.  Synthetic colors from the pure helium atmosphere models
of Bergeron, Wesemael, \& Beauchamp (1995) may be compared.  The v-i
slope is probably least affected by metallic absorption.  The measured
value of +0.24 from our spectrum compares with +0.266 for a 6,000~K log
g=8 atmosphere, and +0.152 at 6,500~K.  From this we may conclude that
LHS~2534 has a $T_{\rm eff}$ near 6,000~K.  The star is clearly warmer
than the heavily-blanketed LP~701-29 (Dahn et al 1978) for which
Kapranidis \& Liebert (1986) estimated $T_{\rm eff} = 4,500$~K.  The v-i
measurement of Greenstein (1984) suggests $\sim$4,800~K from the
pure-helium models.  Both g-r and especially b-v are substantially
redder than the pure-He models predict.  Likewise, LHS~2534 is cooler
than the heavily-blanketed DZ star G~165-7, for which Wehrse \& Liebert
(1980) estimated 7,500~K and the v-i color (Greenstein 1984) suggests
7,100~K.  Perhaps the most similar of the well known DZ stars is
van~Maanen~2, at g-r = +0.26 and v-i = +0.13.  Bergeron, Ruiz \& Leggett
(1997) estimate 6,770~K from fitting a multi-color energy distribution
of this star.

\subsection{The Sodium Triplet and the Magnetic Field Strength}

The Zeeman effect on neutral sodium is a classic problem (Zeeman 1897)
that is encountered here for the first time in regard to white dwarfs.
Thus, we briefly summarize the situation.  Sodium is isoelectronic with
hydrogen, so magnetic effects involve only the single valence electron.
The D$_1$ ($\lambda5895.9$) and D$_2$ ($\lambda5890.0$) lines comprise a
resonance doublet that couples the $^2S_{1/2}$ term with $^2P_{1/2}$ and
$^2P_{3/2}$, respectively.  The feature is also seen in G~165-7 (Hintzen
\& Strittmatter 1974, Wehrse \& Liebert 1980), where the doublet
splitting cannot be resolved due to pressure broadening.

In a weak magnetic field ($B\lesssim10^5$~G), the $J=1/2$ levels are each
split into 2 sublevels, and the $J=3/2$ level splits into 5 according to
$M_J$, and the magnitude of the splitting is computed according to LS coupling.
Ten distinct components result. This is the regime normally encountered in
solar observations (e.g., Beckers 1969; Caccin, Gomez, \& Severino 1993).  The
spin and orbit decouple when the splitting due to the external magnetic field
overwhelms that due to the fine-structure effect.  In Figure~1, the
\ion{Na}{1} feature appears as a strong triplet at $\lambda\lambda$5862, 5892,
and 5924. The observed splitting is not only considerably larger than the
6~\AA\ fine-structure effect, but the pattern is centered near the mean
wavelength of the nonmagnetic doublet. Hence, we conclude that the Paschen-Back
approximation is appropriate and we analyze the feature as an ordinary linear
Zeeman triplet with an insignificant quadratic component.  The displacements
of the $\sigma$ components are then $\Delta(1/\lambda) = +87$~cm$^{-1}$ and
$-$92~cm$^{-1}$ for $\lambda$5862 and $\lambda$5924, respectively.  (It is
customary in atomic spectroscopy to use wavenumber units.)  From the linear
Zeeman effect, the mean surface field (cf. Garstang 1977) is computed
according to
\begin{equation}
B({\rm G}) = {\Delta(1/\lambda) ({\rm cm}^{-1}) \over 4.6686\times10^{-5}}
\end{equation}
which yields an average value for the two components of $B=1.92\times10^6$~G =
1.92~MG. There are, in fact, direct laboratory measurements of the \ion{Na}{1}
D lines which overlap this field strength and corroborate the accuracy of the
linear approximation. Garn et al. (1966) reported splittings between the
$\sigma$ components that from 30~\AA\ for a longitudinal field of 0.94~MG up to
163~\AA\ at 5.1~MG.

Our measurement of 1.92~MG is the mean surface field strength.  Detailed
modeling of the line profile, preferably supported with spectropolarimetric
observations, is necessary to draw conclusions about the field geometry.  A
dipolar geometry is usually an adequate approximation, though the pattern is
often offset significantly from the center of the star.  Time-dependent
observations might determine if the star rotates, and allow the modeling of
periodic changes in the geometric view.  Both $\sigma$ components and,
 to a lesser extent, the $\pi$ component should be
circularly polarized, while linear polarization and polarization of the
continuum should be small.

\subsection{Magnesium and Calcium}

The subordinate \ion{Mg}{1} triplet connects levels $^3S_1$ with $^3P_{0,1,2}$
for $\lambda\lambda$5167.3, 5172.7, 5183.6, respectively.  In LHS~2534 the
region shows 4 principal components at wavelengths of $\sim$5149~\AA,
5180~\AA, 5205~\AA, and 5235~\AA.  Modeling each component of the parent
triplet as a simple Zeeman triplet in a field of $B=1.92$~MG indeed produces a
complex with only 4 lines due to overlapping of some of the 9 components.  The
short-wavelength edge matches that of the data, but the splitting between
lines is somewhat less than observed and thus the feature does not extend
sufficiently far to the red.  We take this as evidence that the linear Zeeman
approximation has broken down for this ion, where the fine-structure effect is
comparable to the magnetic interaction.  We are aware of no computations of
the behavior of \ion{Mg}{1} in this intermediate regime.

The \ion{Ca}{2} ion is isoelectronic with \ion{Na}{1}, but has spin-orbit
splitting which even exceeds that of the \ion{Mg}{1} features and results in
the well-known ``H'' (3933~\AA) and ``K'' (3968~\AA) doublet components being
well-resolved even at low spectral resolution in zero field.  Since the linear
magnetic term in the Hamiltonian is thus comparable to the spin-orbit term,
the splitting at such a low field is more complicated still. Calculations have
been published by Kemic (1975).  
At a field strength of 1.9~MG, the 10 Zeeman components of the doublet
group themselves into 3 features centered around
$\lambda\lambda3925,3954,3987$, and comprised primarily of transitions
from upper levels $^2P_{3/2}$ $\Delta M = -1$, $^2P_{3/2}$ $\Delta M =
0$, and $^2P_{1/2}$ $\Delta M = 0$. These are heavily blended for the
observed line widths, and together result in the broad depression
centered near 3957~\AA. Strong absorption features due to \ion{Fe}{1} and many
other heavy elements are prominent in the spectra of late type stars
shortward of 4000~\AA, and may also contribute in LHS~2534. 
Finally, the strong neutral calcium resonance line at
4226~\AA\ shows a complex structure, and any magnetic components are severely
blended.

\acknowledgments INR would like to thank Fred Chaffee for undertaking the
Keck service observations. We are grateful to D.T. Wickramasinghe for helpful
discussions. JL acknowledges the hospitality of the Institute for Theoretical
Physics, U.C. Santa Barbara, at which part of this work was completed.  The
ITP is supported in part by the National Science Foundation grant PHY
94-07194. GDS thanks the Australian National University for hospitality and
support during a sabbatical leave.
 Support for the study of magnetic stars and stellar systems at
Steward Observatory is also provided by the NSF through grant AST 97-30792 to
GDS.

\clearpage

\begin{figure}
\plotone{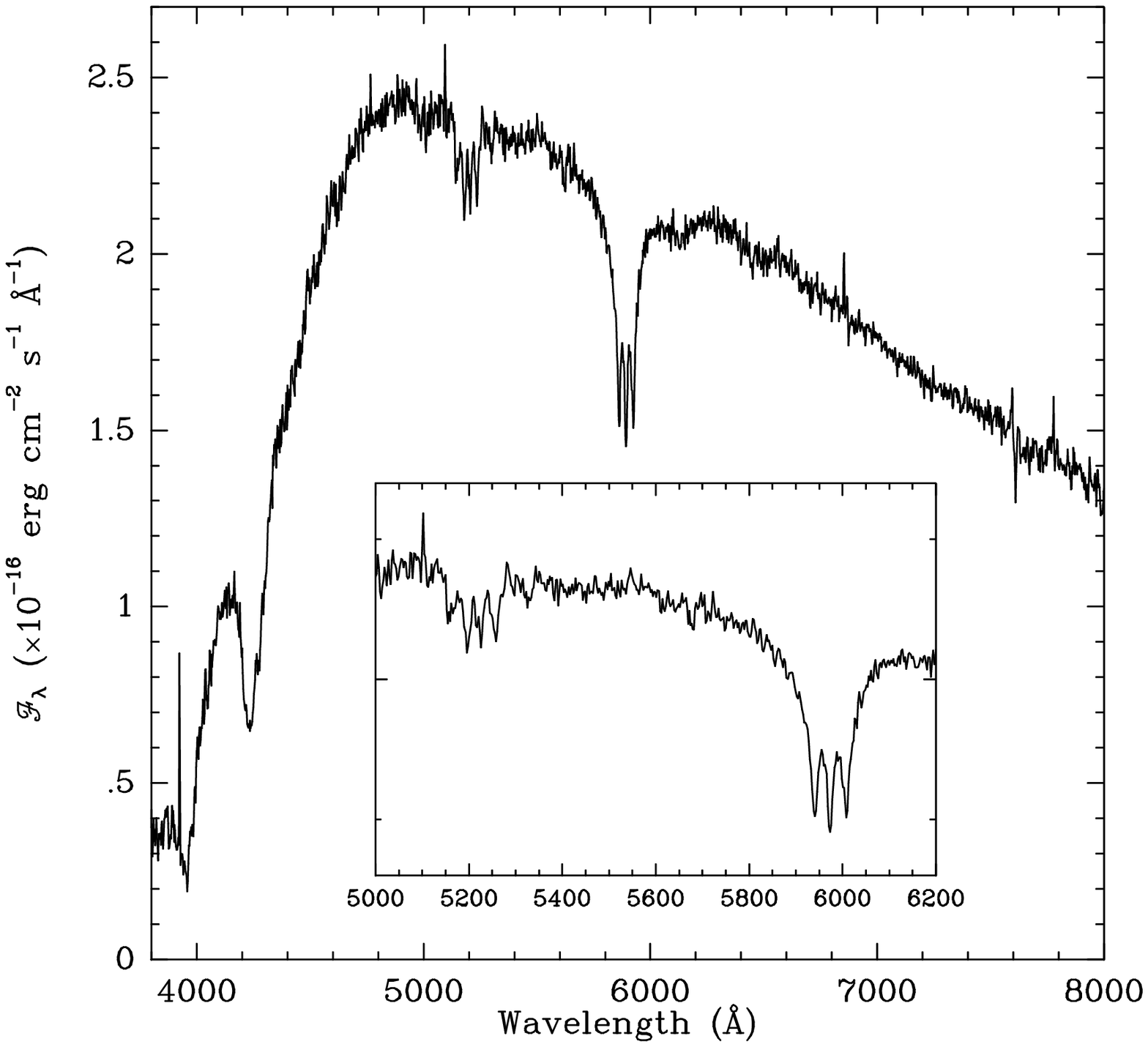}
\caption{The Keck~II / LRIS spectrum of LHS~2534 showing strong \ion{Ca}{2},
\ion{Ca}{1}, \ion{Mg}{1} and \ion{Na}{1} lines.  The clear Zeeman splitting
of the latter two features is examined in the inset.}
\end{figure}

\end{document}